\documentstyle[prl,aps]{revtex}

\begin{document}
\draft

\title{A Self-Consistent Vacuum for Misner Space and\\ the Chronology
Protection Conjecture}
\author{Li-Xin Li and J. Richard Gott, III}
\address{Department of
Astrophysical Sciences, Princeton University, Princeton, NJ 08544}
\date{Received September 5, 1997}
\maketitle

\begin{abstract}
In this paper we find a self-consistent vacuum for Misner
space. For this vacuum the renormalized stress-energy tensor is zero
throughout the Misner space. A point-like particle detector traveling
on a timelike geodesic in a
Misner space with this vacuum detects nothing.  Misner space with 
this vacuum thus creates no problems for time travel in and of 
itself but a time traveler may pose
a danger to himself and to the spacetime.
\end{abstract}
\pacs{PACS number(s): 04.62. +v}

Recently, many people have considered the problem of time travel
\cite{mor88,fro90,got91}. Time travel, i.e. traveling back in time, 
requires the existence
of closed timelike curves (CTCs). In classical general relativity
there are many spacetimes with CTCs solving the Einstein 
equation (for a review see Ref. \cite{tho93}). But, 
in quantum field theory in curved
spacetime, several calculations reveal that the renormalized
stress-energy tensor of the vacuum polarization in a spacetime with
CTCs diverges at the Cauchy
horizon which separates the region with CTCs from that without
closed causal curves, or at the polarized hypersurfaces which are
nested inside the Cauchy horizon
\cite{his82,kim91,fro91,bou92,kli92,gra93}. 
Misner space is a simple example of spacetime 
with CTCs, which is obtained from 
Minkowski space by identifying points that are taken to each other by
the Lorentzian boost transformation \cite{mis67}. Hiscock and
Konkowski have calculated the vacuum polarization of a conformally
coupled scalar field in Misner space and found that the renormalized
stress-energy tensor diverges at the Cauchy horizon \cite{his82}. Based on this
result, Hawking has postulated the chronology protection conjecture
which states that physical laws do not allow the appearance of
CTCs \cite{haw92}. (For counter-examples to this conjecture see
\cite{li94,li96,sus97}.) The vacuum state in Hiscock and
Konkowski's calculation is the usual Minkowski vacuum with multiple images
in the covering space -- Minkowski space. This means that this
``adapted'' Minkowski vacuum (i.e. the Minkowski vacuum with multiple images) 
is not a self-consistent vacuum state for
Misner space because its geometry is locally flat and so the Einstein
equation demands that the
renormalized stress-energy tensor is zero. 
In this paper we find a self-consistent vacuum state
for Misner space. For such a vacuum the renormalized stress-energy
tensor is zero throughout the Misner space. We also discuss the
behavior of a
point-like particle detector in Misner space with such a vacuum and
find no divergent effects on the detector.

Consider
Minkowski space ($R^4$, $\eta_{ab}$) with the metric $\eta_{ab}$
given by (throughout the paper we use the units $c=G=\hbar=1$)
\begin{eqnarray}
   ds^2=-dt^2+dx^2+dy^2+dz^2,
\end{eqnarray}
where $(t,x,y,z)$ are the usual Cartesian coordinates. Making the
transformation
\begin{eqnarray}
   t=\xi\sinh\eta,~~~x=\xi\cosh\eta,
\end{eqnarray}
the metric is transformed to the Rindler metric
\begin{eqnarray}
   ds^2=-\xi^2d\eta^2+d\xi^2+dy^2+dz^2.
\end{eqnarray}
The Rindler coordinates $(\eta,\xi,y,z)$ only cover the right quadrant
(R) of the Minkowski space (i.e. the region $x>|t|$). By 
reflection $(t,x,y,z)\rightarrow(-t,-x,y,z)$ (or
$(\eta,\xi,y,z)\rightarrow$ $(\eta,-\xi,y,z)$), the Rindler
coordinates  
and the Rindler metric
can be extended to the left quadrant (L) ($x<-|t|$).
By the reflection with respect to the hypersurfaces $x=\pm
t$, or
\begin{eqnarray}
   \eta\rightarrow\tilde{\xi}-i{\pi\over2},~~~
   \xi\rightarrow\pm i\tilde{\eta},~~~
   y\rightarrow y,~~~
   z\rightarrow z,
\end{eqnarray}
the Rindler coordinates can be extended to the future quadrant (F)
($t>|x|$) and the past quadrant (P) ($t<-|x|$). The
Rindler metric is extended to F and P as
\begin{eqnarray}
   ds^2=-d\tilde{\eta}^2+\tilde{\eta}^2d\tilde{\xi}^2+dy^2+dz^2.
\end{eqnarray}
Misner space is obtained by identifying $(t,x,y,z)$ with $(t\cosh
nb+x\sinh nb, x\cosh nb+t\sinh nb, y, z)$ where $n$ is any integer and $b$
is a positive boost constant. Under such an identification the point
$(\eta,\xi,y,z)$ in R (or L) is identified with the points
$(\eta+nb,\xi,y,z)$ in R (or L), the point $(\tilde{\eta},
\tilde{\xi},y,z)$ in F (or P) is identified with points  $(\tilde{\eta},
\tilde{\xi}+nb,y,z)$ in F (or P). Clearly there are CTCs
in R and L but there are no CTCs in F and P, and these regions 
are separated by
the Cauchy horizons $x=\pm t$ which are generated by closed null geodesics.

Usually there is no well-defined quantum field theory in spacetimes
with CTCs. However, in the simple case of Misner space, we can do it
in the covering space -- Minkowski space. In Minkowski space there are
two familiar vacuum states: the Minkowski vacuum, which is invariant
under the Poincar$\acute{e}$ group; and the Rindler vacuum, which is not
Poincar$\acute{e}$ invariant but is invariant under the Lorentz
boost transformation \cite{ful73}. It is well known that the
Minkowski vacuum is a thermal state relative to the Rindler vacuum, a
particle detector with constant acceleration moving in the Minkowski
vacuum behaves as if it is in a thermal bath \cite{unr76}. 
Hiscock and Konkowski have calculated the vacuum polarization
of a conformally coupled scalar field in the ``adapted'' Minkowski
vacuum for
Misner space and found that the renormalized stress-energy tensor diverges at
the Cauchy horizon, which has caused Hawking to conclude that such a
spacetime cannot be stable against vacuum fluctuations and
that therefore the chronology protection conjecture holds. 
But, this means that the ``adapted''
Minkowski vacuum is {\em not} a self-consistent
quantum state for the Misner space because the Einstein equation
is not satisfied. 

In the case of an eternal Schwarzschild black
hole, there are the Boulware vacuum \cite{bou75} and the Hartle-Hawking vacuum
\cite{har76}. The globally defined Hartle-Hawking vacuum bears essentially the
same relationship to the Boulware vacuum as the Minkowski vacuum does
to the Rindler vacuum \cite{sci81}. For the Boulware vacuum, the renormalized
stress-energy tensor diverges at the horizon, which means that this
state is not a self-consistent vacuum for the Schwarzschild black hole
because when one inserts this stress-energy tensor back into the
Einstein equation the back-reaction will completely alter the
Schwarzschild geometry near the horizon. For the Hartle-Hawking vacuum
however the renormalized stress-energy tensor is finite everywhere and an
static observer outside the horizon sees Hawking radiation \cite{can80}. People
usually regard the Hartle-Hawking vacuum as the reasonable vacuum
state for an eternal Schwarzschild black hole because when the
stress-energy tensor is inserted back into the Einstein equation the
Schwarzschild geometry is only altered slightly \cite{yor85}. Therefore in the
case of Misner space we should try to find a vacuum which is
self-consistent. Let us consider the Rindler vacuum. The Hadamard
function for a conformally coupling scalar field in the Rindler
vacuum is \cite{dow78}
\begin{eqnarray}
   G_R^{(1)}(X,X^\prime)\equiv\langle 0_R\vert\phi(X)\phi(X^\prime)+
   \phi(X^\prime)\phi(X)\vert 0_R\rangle 
   ={1\over2\pi^2}{\gamma\over\xi\xi^\prime
   \sinh\gamma[-(\eta-\eta^\prime)^2+\gamma^2]},
\end{eqnarray}
where $X=(\eta,\xi,y,z)$,
$X^\prime=(\eta^\prime,\xi^\prime,y^\prime,z^\prime)$ and $\gamma$ is
defined by
\begin{eqnarray}
   \cosh\gamma={\xi^2+{\xi^\prime}^2+(y-y^\prime)^2+(z-
   z^\prime)^2\over2\xi\xi^\prime}.
\end{eqnarray}  
$G_R^{(1)}$ in Eq.(6) is well-defined in region R and can be
analytically extended to region L by $(\eta,\xi,y,z)$ $\rightarrow$
$(\eta,-\xi,y,z)$, and regions F and P by the transformations in
(4). Using the method of images, the corresponding Hadamard function
in Misner space is
\begin{eqnarray}
   G^{(1)}(X,X^\prime)=\sum_{n=-\infty}^{\infty}{1\over2\pi^2}
   {\gamma\over\xi\xi^\prime
   \sinh\gamma[-(\eta-\eta^\prime+nb)^2+\gamma^2]}.
\end{eqnarray}
The Hadamard function for the usual Minkowski vacuum is
\begin{eqnarray}
   G_M^{(1)}(X,X^\prime)={1\over2\pi^2}
   {1\over
   -(t-t^\prime)^2+(x-x^\prime)^2+(y-y^\prime)^2+(z-z^\prime)^2}.
\end{eqnarray}
As usual the regularized Hadamard function is
taken to be
\begin{eqnarray}
   G_{\rm reg}^{(1)}=G^{(1)}-G_M^{(1)}.
\end{eqnarray}
The renormalized stress-energy tensor is obtained from the Hadamard function
by \cite{wal78,bir82}
\begin{eqnarray}
   \langle T_{ab}\rangle_{\rm ren}={1\over2}\lim_{X^\prime\rightarrow
   X}({2\over3}\nabla_a\nabla_{b^\prime}
   -{1\over3}\nabla_a\nabla_b-{1\over6}g_{ab}
   \nabla_c\nabla^{c^\prime})G^{(1)}_{\rm reg}.
\end{eqnarray}
Inserting Eqs.(8-10) into Eq.(11), we obtain the renormalized
stress-energy tensor for the vacuum with the Hadamard function in (8)
\begin{eqnarray}
   \langle T_{ab}\rangle_{\rm
   ren}={1\over1440\pi^2\xi^4}\left[\left({2\pi\over
   b}\right)^4-1\right]
   (4\xi^2d\eta_ad\eta_b+g_{ab})
\end{eqnarray}
This is the renormalized
stress-energy tensor in R. We find that unless 
$b=2\pi$, $\langle T_{ab}\rangle_{\rm
ren}$ blows up as one approaches the Cauchy horizon
($\xi\rightarrow0$). But, for the case $b=2\pi$ we have 
\begin{eqnarray}
   \langle T_{ab}\rangle_{\rm ren}=0,
\end{eqnarray}
which is regular on the Cauchy horizon and can be regularly extended to 
regions L, F, and P where it is also zero. With such a vacuum state (whose Hadamard function is
given by Eq.(8) with $b=2\pi$), the Misner space is self-consistent
i.e. the vacuum Einstein equation is satisfied exactly. Therefore such
a vacuum state is self-consistent and $b=2\pi$
is the {\em self-consistent condition}. (Another solution for $\langle
T_{ab}\rangle_{\rm ren}=0$ in Eq.(12) is $b=i2\pi$ which is just the usual
Minkowski vacuum expressed as a thermal state relative to a Rindler
observer. Periodicity in imaginary time is a characteristic of a thermal
Green function \cite{fet71,gib78}.)

Another way to deal with quantum fields in spacetimes with CTCs is to
do the quantum field theory in the Euclidean section and then
analytically extend the results to the Lorentzian section \cite{haw95}. For Misner
space the Euclidean section is obtained by taking $\eta$ and $b$ to be
$-i\bar{\eta}$ and $-i\bar{b}$. The resultant space is the Euclidean
space with metric $ds^2=\xi^2d\bar{\eta}^2+d\xi^2+dy^2+dz^2$ 
and $(\bar{\eta},\xi,y,z)$ and $(\bar{\eta}+n\bar{b},\xi,y,z)$ are
identified where $(\bar{\eta},\xi,y,z)$ are cylindrical polar
coordinates with $\bar{\eta}$ the angular polar coordinate and $\xi$ the
radial polar coordinate. The geometry at the 
hypersurface $\xi=0$ is conical singular
unless $\bar{b}=2\pi$. When extending that case to the Lorentzian section, we
get $b=2\pi$ which is just the self-consistent condition. This may be
the geometrical explanation of the self-consistent condition. By doing
quantum field theory in the Euclidean space, then analytically
extending the results to the Lorentzian section, we obtain the
renormalized stress-energy tensor in R (or L) region of the Misner
space. Then we can extend the renormalized stress-energy tensor in R (or L) to
regions F (or P). The results are the same as that obtained with
the method of images. 

Now we have a self-consistent vacuum state for the conformally coupled
scalar field in the Misner space
satisfying the self-consistent condition, whose
Hadamard function is given by Eq.(8) with $b=2\pi$. If the quantum field is in
such a state, the renormalized stress-energy 
tensor vanishes everywhere, and the
Einstein equation is satisfied exactly. In order to see if Misner
space in such a quantum state allows time travel, we consider a
point-like particle detector moving in this space. The detector
is assumed to be in its ground state originally. The probability
that the detector is excited is given by the response function \cite{bir82}
\begin{eqnarray}
   {\cal F}(E)=\int_{-\infty}^{\infty}d\tau\int_{-\infty}^\infty 
   d\tau^\prime
   e^{-iE(\tau-\tau^\prime)}G^+(X(\tau),X(\tau^\prime)),
\end{eqnarray}
where $\tau$ is the proper time of the detector, $X(\tau)$ is the
worldline of the detector, $E(>0)$ is the difference of energy between
the excited state and the ground state of the detector,
$G^+(X,X^\prime)$ $\equiv$
$\langle\vert\phi(X)\phi(X^\prime)\vert\rangle$ is the Wightman
function.
Suppose the detector moves along a geodesic with $x=a$, $y=\beta_yt$, 
and $z=0$ ($a$ and $\beta_y$ are constants and $a$ is positive), which 
goes through the P, R, and F regions. The proper time of the detector is
$\tau=t/\zeta$ with $\zeta=1/\sqrt{1-\beta_y^2}$. Along
the geodesic, the Hadamard function in (8) is reduced to
\begin{eqnarray}
   G^{(1)}(t,t^\prime)={1\over2\pi^2}{\gamma\over
   \sinh\gamma\sqrt{(a^2-t^2)(a^2-{t^\prime}^2)}}
   \sum
   _{n=-\infty}^{\infty}{1\over-(\eta-\eta^\prime+nb)^2+\gamma^2},
\end{eqnarray}
where $\gamma$ is given by
\begin{eqnarray}
   \cosh\gamma={2a^2-t^2-{t^\prime}^2+\beta_y^2(t-t^\prime)^2\over
   2\sqrt{(a^2-t^2)(a^2-{t^\prime}^2)}},
\end{eqnarray}
and $\eta-\eta^\prime$ is given by
\begin{eqnarray}
   \sinh(\eta-\eta^\prime)={a(t-t^\prime)\over
   \sqrt{(a^2-t^2)(a^2-{t^\prime}^2)}}.
\end{eqnarray}
Though this Hadamard function is originally defined only in R it can be
analytically extended to F, P, and L. The Wightman function is equal to
$1/2$ of the Hadamard function with $t$ replaced by
$t-i\epsilon/2$ and $t^\prime$ replaced by
$t^\prime+i\epsilon/2$  
where $\epsilon$ is an infinitesimal positive
real number. Then the response function is
\begin{eqnarray}
   &&{\cal F}(E)={1\over4\pi^2}\sum_{n=-\infty}^{\infty}
   \int_{-\infty}^{\infty}dT\int_{-\infty}^\infty d\Delta\tau \nonumber\\
   &&{\gamma^+e^{-iE\Delta\tau}\over\sinh\gamma^+\sqrt{[a^2-\zeta^2
   (T+{\Delta\tau\over2}-{i\epsilon\over2\zeta})^2]
   [a^2-\zeta^2(T-{\Delta\tau\over2}+{i\epsilon\over2\zeta})^2]}\{-
   [(\eta-\eta^\prime)^++nb]^2+{\gamma^+}^2\}},
\end{eqnarray}
where $T\equiv(\tau+\tau^\prime)/2$,
$\Delta\tau\equiv\tau-\tau^\prime$; 
$\gamma^+$ and $(\eta-\eta^\prime)^+$ are given by (16) and (17)
with $t$ replaced by
$t-i\epsilon/2$ and $t^\prime$ replaced by
$t^\prime+i\epsilon/2$. The
integral over $\Delta\tau$ can be worked out by the residue theorem where
we choose the integration contour to close in the lower half
complex-$\Delta\tau$ plane. The result is zero since there are no poles
in the lower half plane. Therefore such a detector
cannot be excited and so it detects nothing. It is interesting
to note that
such a conclusion holds even for a simply connected Minkowski space:
An inertial particle detector moving in the Rindler vacuum detects
nothing though the Rindler vacuum is different from the Minkowski
vacuum. This can be explained by the fact that the energy level of the Rindler
vacuum is below that of the Minkowski vacuum \cite{can77} therefore the Rindler
vacuum cannot excite an inertial detector in its ground state. 
We have also calculated the response
functions for detectors on worldlines with constant $\xi$, $y$,
and $z$ and worldlines with constant $\tilde{\xi}$, $y$,
and $z$, both are zero.

However, in this particular spacetime a time traveler may pose a
danger to himself - and to the spacetime as a whole. The time traveler
will collide with himself unless his worldline is
carefully arranged. A time traveler moving along the geodesic
discussed above will hit himself unless $\beta_y$ is non-zero and
his dimension is sufficiently
small compared with $a$. In the 
covering space an infinite number of
images of the time traveler must pass inside the sphere
$x^2+y^2+z^2=a^2$ at $t=0$. So unless the total mass of the time
traveler and his spaceship is {\em exactly zero} it 
will distort the geometry of the
spacetime and the exact solution will not apply. This may be avoided
if the time traveler and his spaceship are arranged to have a
spherically symmetric mass distribution and are surrounded by a
negative mass shell so that the total mass is zero and the
gravitational field outside the shell is zero. 
(This would require violating the weak energy
condition as is done in any case in the wormhole solution
\cite{mor88}). The time traveler must also not emit any 
stray photons that he does not later re-absorb.

Misner space suffers from classical instability in the sense
that a propagating wave may pile up on the Cauchy horizon and
destroy the Cauchy horizon due to the infinite blue-shift of the
frequency of the wave \cite{mor88}. This is similar in nature to the
perturbation on Misner space caused by a positive mass time traveler discussed
above. However, this classical instability can be avoided in the
wormhole spacetime with CTCs
due to the divergent effect of the wormhole \cite{mor88}. (The
relation between Misner space and the wormhole spacetime with CTCs
can be found in \cite{tho93}).

Grant space is identical to Misner space except that in Grant
space there is a
shift in the $y$ direction when one makes the identification in
Minkowski space \cite{gra93}. Grant space is identical to Gott's
two-string spacetime far from the strings \cite{gra93,lau94} 
but different in that Gott
space has some Casimir effects that Grant space does not. In Grant
space the problem of the time traveler 
hitting himself (e.g. for a time traveler with
$x=a$ and $y=z=0$) is avoided if the
dimension of the time traveler is less than the
shift in the $y$ direction. But Thorne has conjectured that black holes
with horizons form when and only when a mass $M$ gets compacted into a
region whose circumference in every direction is ${\cal C}<4\pi M$
\cite{tho72}. At $t=0$, the pair of $n$-th and $(-n)$-th images of the time
traveler form a system with total mass $M=2m\cosh nb$ in their
center of mass frame, where $m$ is the rest mass of the
time traveler. The circumference is ${\cal C}=4ny_0$ where $y_0$ is the
shift in the $y$ direction. If $m>0$ one can always find an $n$ large
enough so that $4ny_0<4\pi\cdot2m\cosh nb$ and ${\cal C}<4\pi M$
thus causing a black hole to form according to Thorne's 
conjecture. Then the spacetime
would be significantly perturbed perhaps eliminating the
CTCs. Therefore Grant space appears to be classically unstable to a
perturbation by a positive mass time traveler. And so also 
by implication is Gott's
two-string spacetime with CTCs. For the finite string loop case,
Gott's two-string spacetime with CTCs already satisfies Thorne's
limit for black hole formation, suggesting that the CTCs would either
be destroyed or hidden behind an event horizon \cite{got91,hea94}. 
We have found a self-consistent vacuum for
Grant space only in the case of $y_0=0$, but in such a case Grant
space is reduced to Misner space. 

We have also found a self-consistent vacuum for the multiply connected de
Sitter space which is obtained from the usual de Sitter space by
identifying points that are taken to each other by the de Sitter-boost
transformation, which will be discussed elsewhere \cite{got97}. In
such a case the renormalized stress-energy tensor is non-zero but the
Einstein equation is exactly solved for some specific cosmological
constant (as in \cite{got82}).

\acknowledgments
This research was supported by NSF grant AST95-29120.

~~~

{\em  Note added} (11/24/97):
It has just come to our attention that M. J. Cassidy (Class. Quantum
Grav. {\bf 14}, 3031 (November 1997)) has independently argued, using
Euclidean methods and starting from the vacuum around a cosmic string,
that a solution with $\langle T_{ab}\rangle_{\rm ren}=0$ must exist
for a conformal scalar field in Misner space where $b=2\pi$ (in
agreement with our result). He didn't propose what quantum state it
would correspond to (we have found here that it is an ``adapted''
Rindler vacuum), but he noted correctly that it would have to be
different than the Hiscock and Konkowski vacuum. We agree with
C. R. Cramer and B. S. Kay's (Class. Quantum Grav. {\bf 13}, L143
(1996)) conclusion (in the automorphic field \cite{sus97} case) that
$\langle T_{ab}\rangle_{\rm ren}$ remains formally ill-defined on the
Misner space Cauchy horizon itself, but we note that this may not be a
problem in physics since this is a set of measure zero, and $\langle
T_{ab}\rangle_{\rm ren}=0$ everywhere else.

\end{document}